# Wetting-induced budding of vesicles in contact with several aqueous phases

*Yanhong Li[†], Halim Kusumaatmaja[‡], Reinhard Lipowsky, and Rumiana Dimova\**

Max Planck Institute of Colloids and Interfaces, Science Park Golm, 14424 Potsdam, Germany.

ABSTRACT: Osmotic deflation of vesicles enclosing two liquid phases can lead to bulging of one of the phases from the vesicle body. This budding process is preceded by a complete to partial wetting transition of one of the liquid phases on the membrane and depends on the membrane tensions and the tension of the interface between the enclosed liquid phases. These tensions dominate in different morphology regimes, the crossover of which initiates the budding process. In addition, the degree of budding can be controlled by aspiration via micropipettes. We also demonstrate that the budding direction can be reversed if there are two external phases in contact with the vesicle.

1. INTRODUCTION.

Budding processes of biological membranes are essential for cell functioning. They are involved, for example, in vesicular trafficking, which requires the selective loading of cargo molecules into the vesicles, or in endo- and exocytosis, where the membrane invaginates inward or outward. In giant vesicles, model membrane systems with cell-size dimensions[1], budding may be induced by a number of possible mechanisms. For example, slight area difference between the two monolayer leaflets of the membrane or a change in the vesicle area-to-volume ratio may lead to budding[2,3]. In homogeneous membranes, budding is driven by the minimization of the bending energy[4,5]. In multicomponent membranes containing phase separated domains, budding is governed by the interplay of bending rigidity and line tension of these domains[6-9].

In the aforementioned studies, the membrane properties play a crucial role, whereas the enclosed homogeneous fluid is an inert "spectator" phase. However, when two (or more) liquid phases are in contact with the membrane, their interactions with the bilayer become relevant for the vesicle morphology. Recently, such novel soft matter systems were realized[10-13] and phase separation inside vesicles was reported to lead to protein partitioning in one of the liquid phases[10,11]. Budding has also been observed, but the underlying mechanism remains to be elucidated.

To shed light on this problem, we use giant unilamellar vesicles loaded with homogeneous aqueous solutions of water soluble polymers. Osmotic deflation of the vesicle induces phase separation in the interior solution and, at the same time, creates excess membrane area. Depending, among others, on the membrane tension and spontaneous curvature, the area created during deflation may be stored in membrane nanotubes[13] and/or may lead to vesicle budding as shown in this paper. The budding direction can be reversed if the phase separation occurs in the vesicle exterior. Systematic study of the budding process reveals that it requires the wetting of the membrane by both liquid phases. This wetting-induced budding provides a possible mechanism for the selective loading of cargo molecules into transport vesicles in cells.

## 2. EXPERIMENT.

To prepare vesicles encapsulating two phases we used homogeneous aqueous solution composed of 4.05 wt % poly(ethylene glycol) (PEG, 8 kg/mol) and 2.22 wt % dextran (400-500 kg/mol), see Supporting Information. The membrane consists of 95.9 mol % dioleoylphosphatidylcholine, 4.0 mol % $G_{M1}$ Ganglioside and 0.1 mol % fluorescent lipid marker, see Supporting Information. After formation, the vesicles are diluted in an isotonic solution containing 4.41 wt % PEG and 1.45 wt % dextran. The osmolarity of the external medium is then increased by adding a hypertonic solution of 3.92 wt % PEG, 2.14 wt % dextran and 3.27 wt % sucrose. The choice for this solution composition is defined by the requirement that its density is between those of the PEG-rich and dextran-rich phases, but smaller than the overall density of the solution inside the vesicles. These conditions ensure that the vesicles sediment to the chamber bottom and that the symmetry axis of the budded vesicles is vertically oriented; see Fig. 1(a-f). In this way, the vesicles can be observed from the side using a horizontally aligned microscope. The vesicle axial symmetry allows for quantitative analysis of their shapes in contrast to the systems explored previously[10,11].

For vesicle preparation where the phase separation occurs outside the vesicles as in Fig. 1(h-q), we prepared a polymer solution composed of 6.87 wt% PEG and 2.86 wt% dextran. A small fraction of the dextran (1.25 wt% of the total dextran) was labeled with fluorescein isothiocyanate. The polymer solution was left at room temperature (23° C) for 2.5 days after which complete phase separation was achieved. Then, the two phases were separated. Vesicles with $G_{M1}$-doped membrane were prepared in the PEG-rich phase at room temperature using electroformation. Approximately 2 μl dextran-rich phase was mixed with 200 μl vesicle solution in the microscopy chamber (occasionally a mixture of 100 μl PEG-rich phase and 100 μl vesicle solution was used). The chamber was shaken gently in order to break the dextran-rich phase into tiny droplets, and to increase the possibility for contact between the vesicles and the dextran-rich droplets. The vesicles with dextran-rich droplets were imaged after 2 hours or on the next day.

For the micropipette aspiration experiments, performed only on vesicles encapsulating two phases, the vesicles are pre-deflated with the hypertonic solution, increasing the external osmolarity by 32 %, and then, left to equilibrate for one day. Aspiration is realized by means of a glass capillary, approximately 20 μm in diameter, connected to a home-built hydrostatic pressure system. After each consecutive pressure change, the vesicle is left to equilibrate for about 3 minutes.

The interfacial tension of the phase separated polymer solution is independently measured in bulk using spinning drop tensiometer, see Supporting Information. All experiments are performed at room temperature.

3. RESULTS.

*Budding*. Figure 1(a-g) shows side-view images and a cartoon of a vesicle at different degrees of deflation described by the osmolarity ratio $r$, which we define to be the ratio $r_{ex}/r_{in,0}$ between the deflation-dependent osmolarity $r_{ex}$ within the exterior compartment and the initial osmolarity $r_{in,0}$ within the interior compartment. Initially, the vesicle interior is homogeneous; see Fig. 1(a). The phase separation within the vesicle is induced via osmotic deflation by adding the hypertonic solution to the external medium. The resulting osmotic unbalance forces water out of the vesicle. Since the polymers cannot permeate the membrane, their concentration in the vesicle increases. Once it is above the binodal (see Fig. S1 in the Supporting Information), the enclosed solution separates into PEG-rich and dextran-rich phases; see Fig. 1(b). The dextran-rich droplet is always located at the bottom of the vesicle because of its higher density compared to the PEG-rich phase. When the vesicle is further deflated with $r > 1.3$, the dextran-rich phase starts to wet the membrane; see Fig. 1(c). The change in the morphology of the dextran-rich droplet indicates a wetting transition from complete wetting of the PEG-rich phase in Fig. 1(b) to partial wetting in Fig. 1(c) as previously described in Ref. [12].

As the dextran droplet further wets the membrane, the radius $R_c$ of the three-phase contact line, where the external medium (*e*), the PEG-rich phase (*p*) and the dextran-rich phase (*d*) are in close proximity, increases; see Figs. 1(g) and 2(a, b). At the same time, the interfacial tension $\Sigma_{pd}$ between the liquid phases pulls on the membrane along this contact line and leads to budding and a decrease in $R_c$ at higher polymer concentrations; see Fig. 2(b). The budding process from the spherical vesicle shape in Fig. 1(c) to the pair of two spherical caps in Fig. 1(d) should be regarded as a strongly concentration-dependent but smooth crossover rather than as a sharp transition. This crossover behavior is most clearly seen in the behavior of the contact line radius $R_c$, see Fig. 2(b). As the osmolarity is increased further, the dextran-rich phase protrudes further out; see Fig. 1(e, f). Finally, the dextran-rich phase may form a completely spherical bud whereby the area of the *pd* interface is close to zero.

The excess membrane area governs the shape of the budded vesicle. Thus, vesicles with different initial area adopt diverse shapes upon deflation, see Supporting Information. Wetting-induced budding can also occur towards the vesicle interior, see Fig. 1(h-q), when the exterior solution undergoes phase separation. The observations in these samples show analogous vesicle morphologies. For low excess membrane area, the dextran-rich droplets wet the membrane but deform the vesicle only slightly, see Fig. 1(h-j), while much deflated vesicles may engulf the dextran-rich droplets completely, see Fig. 1(n-p).

*Force balance at the three-phase contact line*. After the wetting transition, the surface of the dextran-rich droplet exhibits a "kink" at the three-phase contact line, see Fig. 1(c-h, j, k, m, q). This contact line divides the vesicle membrane into *pe* and *de* segments. Because of the compositional difference of the *p* and *d* phases, these two membrane segments experience two distinct tensions, $\hat{\Sigma}_{pe}$ and $\hat{\Sigma}_{de}$, see Fig. 2(a).

By fitting the vesicle and the drop contours in the microscopy images with circular arcs, we obtain the three contact angles $\theta_e$, $\theta_p$ and $\theta_d$ with $\theta_e + \theta_p + \theta_d = 2\pi$, see Fig. 2(a). For outward budding, these contact angles are related to the membrane tensions $\hat{\Sigma}_{pe}$ and $\hat{\Sigma}_{de}$ and the interfacial tension $\Sigma_{pd}$, via

$$\Sigma_{pd} + \hat{\Sigma}_{pe} \cos\theta_p + \hat{\Sigma}_{de} \cos\theta_d = 0,$$
$$\Sigma_{pd} \cos\theta_p + \hat{\Sigma}_{pe} + \hat{\Sigma}_{de} \cos\theta_e = 0, \quad (1)$$
$$\Sigma_{pd} \cos\theta_d + \hat{\Sigma}_{pe} \cos\theta_e + \hat{\Sigma}_{de} = 0.$$

The case of inward budding is described in the Supporting Information. These equations are analogous to Neumann's triangle for capillary surfaces[14]. From the analysis of the vesicle images, we can calculate the total polymer concentration within the vesicle at every osmolarity ratio $r$, assuming that the volume change corresponds to the loss of water. The interfacial tension $\Sigma_{pd}$ in Eq. (1) is measured independently for bulk polymer solutions, see Supporting Information. Knowing $\Sigma_{pd}$, we can compute the tensions $\hat{\Sigma}_{pe}$ and $\hat{\Sigma}_{de}$ from Eq. (1), see Fig. 2(c). The interfacial tension $\Sigma_{pd}$ increases with the total polymer concentration because the composition difference between the coexisting phases is enhanced. Interestingly, the tensions $\hat{\Sigma}_{pe}$ and $\hat{\Sigma}_{de}$ behave significantly differently from $\Sigma_{pd}$. Their non-monotonic dependence on the polymer concentration arises from two competing effects of the deflation process. First, the total vesicle volume is decreased, which in the absence of any other constraint, would act to reduce the membrane tensions $\hat{\Sigma}_{pe}$ and $\hat{\Sigma}_{de}$. However, deflation also increases the interfacial tension $\Sigma_{pd}$, see Fig. 2(c), which may be balanced in two ways following (1): (*i*) by a reduction of the area of the *pd* interface, which as a consequence leads to a reduction in the contact angle $\theta_e$, or (*ii*) by an increase in the membrane tensions $\hat{\Sigma}_{pe}$ and/or $\hat{\Sigma}_{de}$. Inspection of Fig. 2(c) shows that $\hat{\Sigma}_{pe}$ and $\hat{\Sigma}_{de}$ first decrease at low polymer concentrations, which implies that the increased interfacial tension $\Sigma_{pd}$ can be balanced by a reduction of the contact angle $\theta_e$ alone. However, as the polymer concentration is further increased, the tension $\hat{\Sigma}_{de}$ passes through a minimum and starts to increase whereas the tension $\hat{\Sigma}_{pe}$ stays essentially constant. Thus, in this second concentration range, the increase in the interfacial tension $\Sigma_{pd}$ is primarily balanced by an increase in the tension $\hat{\Sigma}_{de}$. Indeed, we will argue further below that the *pe* membrane segment cannot pull "against" the *pd* interface. The two concentration ranges in which either the membrane tensions or the interfacial tension dominate correspond to two different vesicle morphologies as observed optically: sphere or a pair of spherical caps, see Fig. 1. The crossover between these two morphologies marks the budding process.

*Vesicle micropipette aspiration.* For a fixed composition of the polymer phases, the interfacial tension $\Sigma_{pd}$ is constant. However, the membrane tensions $\hat{\Sigma}_{pe}$ and $\hat{\Sigma}_{de}$ can be externally manipulated. Using a micropipette, we apply suction pressure to a deflated dumbbell-like vesicle, see Fig. 3(a). The pressure produces mild membrane tensions in the range 4 - 383 µN/m, see Supporting Information, which do not lead to membrane stretching[15]. The angle $\theta_e$, which characterizes the degree of budding, increases with the applied suction pressure, i.e., the bud is retrieved and the budding process is reversed. Simultaneously, the excess area of the vesicle part outside of the pipette is reduced, which from morphology viewpoint is analogous to vesicle inflation, i.e., the reverse of the process shown in Fig. 1.

The tension $\hat{\Sigma}_{pe}$ is imposed by the micropipette, see Supporting Information, and the tension $\hat{\Sigma}_{de}$ can be calculated from the measured contact angles via (1). The two tensions normalized by the interfacial tension $\Sigma_{pd}$ are shown in Fig. 3(b) as a function of the angle $\theta_e$. For small $\theta_e$ or high degree of budding, the tensions $\hat{\Sigma}_{pe}$ and $\hat{\Sigma}_{de}$ change only weakly. In the regime of large $\theta_e$ or low degree of budding, these tensions increase significantly. The crossover between these two regimes indicates the budding (or bud retrieval) process, which is continuous.

The sharp increase in $\hat{\Sigma}_{pe}$ and $\hat{\Sigma}_{de}$ at high aspiration pressures or large $\theta_e$ can be understood by considering the force balance at the three-phase contact line as in (1), which imply $\hat{\Sigma}_{pe}/\sin\theta_d = \hat{\Sigma}_{de}/\sin\theta_p = \Sigma_{pd}/\sin\theta_e$. As the contact angle $\theta_e$ approaches $\pi$, small changes in $\theta_e$ lead to significant changes in $\hat{\Sigma}_{pe}$ and $\hat{\Sigma}_{de}$. For the regime of lower aspiration pressures (smaller $\theta_e$), where the changes in the tensions $\hat{\Sigma}_{pe}$ and $\hat{\Sigma}_{de}$ are insignificant, see Fig. 3(b), the vesicle shape is governed primarily by the minimization of the interfacial area $A_{pd}$, see the sequence of images 4 to 1 in Fig. 3(a) and Fig. S4 in the Supporting Information.

In these experiments, the composition of the two phases and the membrane do not change. Thus, the wettability of one of the phases, for example the PEG-rich phase, on the membrane should be constant. If we consider a planar membrane, i.e. $\theta_e = \pi$, this wettability will be characterized by $\cos \theta_p$. To examine the limit of a planar membrane, we plot $\cos \theta_p$ as a function of the angle $\theta_e$ in Fig. 3(c). The planar membrane limit is described by the angle $\theta_p^* \equiv \theta_p (\theta_e \approx \pi)$. Linear extrapolation of the data in the range $155° < \theta_e < 178°$ leads to $\cos \theta_p^* \cong 0.87$ or an angle of 30°. For such a planar surface, the wettability is defined as $(\hat{\Sigma}_{de} - \hat{\Sigma}_{pe})/\Sigma_{pd}$. Following (1), one obtains

$$(\hat{\Sigma}_{de} - \hat{\Sigma}_{pe})/\Sigma_{pd} = (\sin \theta_p - \sin \theta_d)/\sin \theta_e. \quad (2)$$

We will now consider the limit of a planar membrane in which the contact angle $\theta_e$ approaches $\pi$. We expand the right hand side of Eq. (2) with respect to $(\pi - \theta_e)$, see the Supporting Information, which leads to

$$(\hat{\Sigma}_{de} - \hat{\Sigma}_{pe})/\Sigma_{pd} = \cos \theta_p^* + O(\pi - \theta_e). \quad (3)$$

where $O(\pi - \theta_e)$ indicates terms of order $\pi - \theta_e$.

Furthermore, from the experimental data we find that $(\hat{\Sigma}_{de} - \hat{\Sigma}_{pe})/\Sigma_{pd}$ is approximately constant over the whole $\theta_e$ range with an average value of $0.86 \pm 0.02$, see Fig. 3(c), which implies the relation:

$$(\hat{\Sigma}_{de} - \hat{\Sigma}_{pe})/\Sigma_{pd} \cong \cos \theta_p^*. \quad (4)$$

This observation can be understood if one considers the membrane shape on different length scales. When viewed with optical resolution as in Fig. 1 and Fig. 3, the membrane shape exhibits a kink that defines the contact angle $\theta_e$. However, this kink cannot persist to small scales since the membrane has a finite bending rigidity and a sharp kink would imply an infinite bending energy. Therefore, the membrane should be smoothly curved on small scales and its contact with the *pd* interface should be characterized by an intrinsic contact angle $\theta_{in}$, which represents the angle between the *pd* interface and the plane tangent to the membrane along the contact line[16]. The relations (2) and (4) then imply that the angle $\theta_p{}^*$ is identical with the intrinsic contact angle $\theta_{in}$ as introduced in Ref. [16]. It represents the actual contact angle between the membrane and the *pd* interface at nanometer length scale. For the polymer composition as studied in the aspiration experiments, we obtain $\cos\theta_p{}^* \cong 0.87$ from linear extrapolation of $\cos\theta_p$ as in Fig. 3(c) and $\cos\theta_p{}^* = 0.86 \pm 0.02$ from relation (4). Thus, both estimates lead to $\theta_{in} = \theta_p{}^* \cong 0.53$ or 30° providing us with two approaches for measuring the intrinsic contact angle.

Since $\theta_{in}$ is smaller than 90 degrees, the *pe* membrane segment and the *pd* interface pull roughly in the same direction. An increase in the interfacial tension arising, e.g., from an increase in the polymer concentrations, must then be primarily balanced by an increase in the mechanical tension $\hat{\Sigma}_{de}$ of the *de* membrane segment, which agrees with the data in Fig. 2(c) for large polymer concentrations.

## 4. DISCUSSION.

Above, the budding process was analyzed using the force balance relations in Eq. (1), which can be directly applied to the images as obtained by optical microscopy. As explained in reference [16], these force balance relations are valid in the tension-dominated regime, i.e., on length scales that exceed the characteristic scale $\Sigma_{pd}/\kappa \cong 100$ nm, where $\kappa$ is the bending rigidity of the membrane. On smaller scales, one obtains more complex force and torque balance equations, which also depend on the bending rigidity $\kappa$ [16]. However, the bending rigidity does not affect the intrinsic contact angle. Indeed, as shown here, we are able to deduce the existence of this contact angle from the coarse-grained force balance equations (1) alone by considering the limit of a planar membrane, for which the effective contact angle $\theta_e$ approaches the limiting value $\theta_e = \pi$, see the line of arguments after Eq. (2).

Wetting-induced budding requires contact of the membrane with both phases, whereby the budding direction depends on whether the two phases are inside or outside the vesicle, see Fig. 1. If the membrane is wetted completely by one of them, the wetting-induced budding cannot occur. Instead, the excess area created during deflation is stored in nanotubes stabilized by the membrane spontaneous curvature.[13] Therefore, the wetting transition as in Fig. 1(b-c) precedes the budding process.

More generally, the degree of budding and tubular formation results from the competition between the spontaneous curvature of the membrane and the wetting properties of the aqueous phases on the membrane. Spontaneous curvature favors tube formation. Since this process consumes excess membrane area, budding in return becomes hindered. Budding, on the other hand, is favored when neither one of the aqueous phases dominantly wets the membrane (i.e. partial wetting). Since the membrane must be wetted by both aqueous phases during the budding process, it cannot occur if the system exhibits complete wetting.

After the wetting transition, one has $\Sigma_{pd} = (\hat{\Sigma}_{de} - \hat{\Sigma}_{pe})/\cos\theta_{in} > \hat{\Sigma}_{de} - \hat{\Sigma}_{pe}$ corresponding to partial wetting with intrinsic contact angle $\theta_{in}$. The energy cost of increasing the contact area between the membrane and the dextran-rich phase is smaller than the energy gain of decreasing the *pd* interfacial area. With deflation, the tensions $\hat{\Sigma}_{pe}$ and $\hat{\Sigma}_{de}$ decrease and budding occurs spontaneously provided excess membrane area is available. Upon further deflation, the tension $\hat{\Sigma}_{pe}$, which is dominated by the membrane spontaneous curvature[13] saturates, see Fig. 2(c). This behavior reflects the fact that at high polymer concentration the fraction of dextran in the PEG-rich phase becomes negligible, see the Supporting Information.

Wetting-induced budding is not limited to equilibrated vesicles. It can also occur during the phase separation process if the transient tensions satisfy $\Sigma_{pd} > \hat{\Sigma}_{de} - \hat{\Sigma}_{pe}$ and there is excess membrane area. More than one bud can form in a metastable state, as observed in our own experiments and in Ref. [11].

5. CONCLUSIONS.

Vesicles in contact with two phases in the interior or exterior can undergo budding, provided the membrane is partially wetted by both phases and excess area is available. If one of the phases does not wet the membrane, it will exist as a small droplet inside or outside the vesicle. The budding process is governed by the interplay between the membrane tensions $\hat{\Sigma}_{pe}$ and $\hat{\Sigma}_{de}$ and the interfacial tension $\Sigma_{pd}$.

We demonstrated that for a fixed composition of the liquid phases the budding process can be reversed by applying tension to the membrane.

The process of wetting-induced budding provides a possible mechanism for the selective vesicular transport in cells. Molecular crowding widely exists in the interior of cells and can lead to microcompartmentation. The membrane prefers to enclose macromolecular solutions, with which it has higher affinity, thus loading the appropriate molecules in transport vesicles. Wetting of a membrane by a liquid droplet creates a contact line, which may enhance endocytotic or exocytotic processes. Wetting-induced budding of vesicles or cells could be used to locally regulate and compartmentalize cellular processes such as protein synthesis, to restrict chemical reactions to particular segments of the membrane surface or parts of the cell, or to terminate such reactions by dewetting or bud retraction. Wetting-induced budding may also provide simple physical mechanism for vesicle shedding[17], involved in membrane traffic and molecule transfer among neighboring cells or to the cell exterior.

**Supporting Information Available**: Materials; Giant vesicle preparation using electroformation; Phase diagram of the polymer aqueous solution; Batch measurements on deflated vesicles; Force balance and intrinsic contact angle for inward budding; Interfacial tension measurements; Vesicle aspiration with micropipettes; Derivation of Eq. (3) in the main text; Calculated vesicle shapes; Saturation of the tension $\hat{\Sigma}_{pe}$ at high osmolarity ratios.

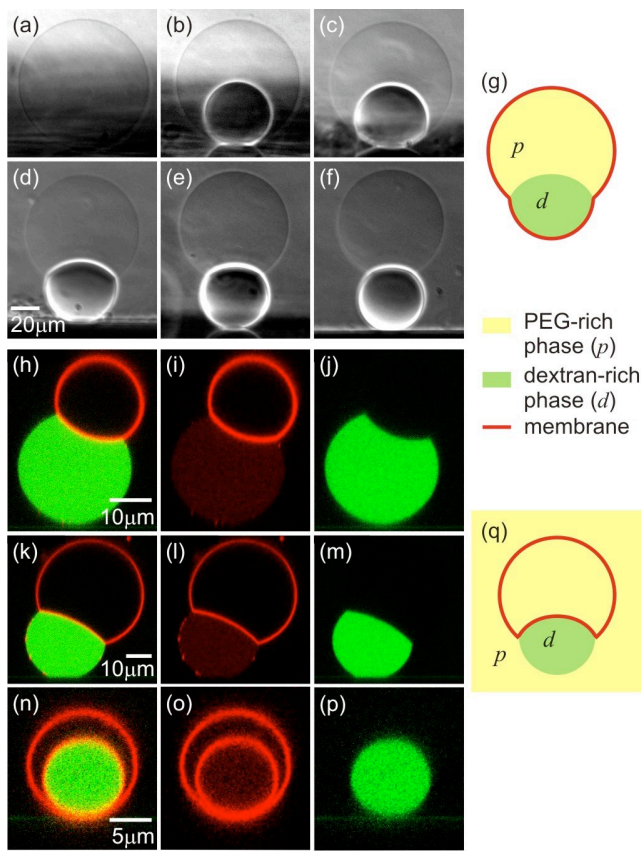

Figure 1. Outward (a-g) and inward (h-q) budding of vesicles in contact with several aqueous solutions. (a-f) Side-view phase contrast images and (g) a cartoon of a vesicle sitting on a glass substrate and subjected to osmotic deflation. Initially, the vesicle contains a homogeneous polymer solution (a). After phase separation (b), the heavier and denser dextran-rich phase is located at the vesicle bottom. Further deflation causes the dextran-rich phase to wet the membrane (c), and then to bud out (d-f). The osmolarity ratio $r$ from (a) to (f) is 1.0, 1.14, 1.32, 1.46, 1.65 and 1.76, respectively. The system was left to equilibrate for at least 2 hours after each consecutive deflation. (h-m) $xz$-confocal scans and (q) a cartoon of vesicles in contact with fluorescently labeled droplets of dextran-rich phase (green). The membrane is also fluorescently labeled (red). The confocal images in the rows (h-j), (k-m) and (n-p) show the mixed fluorescence signal, and the red and green channels separately. Prior to contact with the vesicles, the dextran rich droplets are typically very little or not coated with lipid, see (i, l).

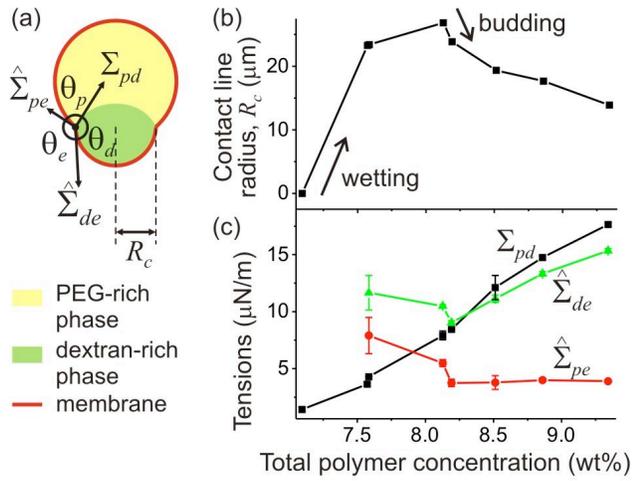

Figure 2. (a) Schematic illustration of the contact angles and the tensions at the three-phase contact line (⊙) with radius $R_c$. (b) Evolution of the radius $R_c$ with increasing polymer concentration. When the dextran-rich phase starts wetting the membrane, $R_c$ first increases, but then decreases during the budding process. (c) Evolution of the three tensions with increasing polymer concentrations. The weight ratio of dextran and PEG is 0.55. The error bars of $\hat{\Sigma}_{pe}$ and $\hat{\Sigma}_{de}$ indicate the error introduced by image fitting. The 1$^{st}$, 3$^{rd}$, 5$^{th}$ and 6$^{th}$ points from these data sets correspond to the images (c-f) in Fig. 1.

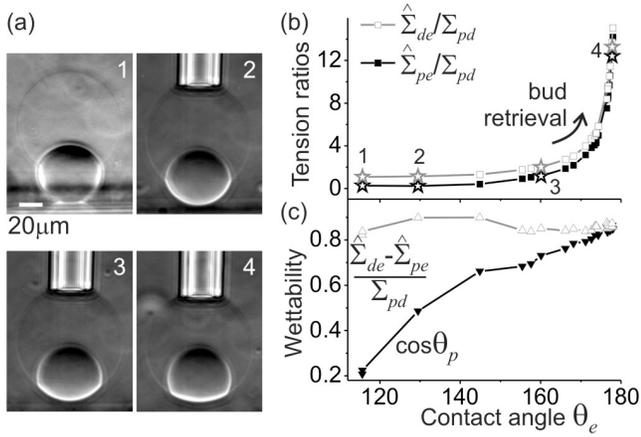

Figure 3. (a) Side-view phase contrast images of a vesicle under different suction pressures. The small dense droplet at the vesicle bottom is the dextran-rich phase. The bright cylindrical part in the second image is the glass micropipette. (b) The tension ratios as a function of the contact angle $\theta_e$ describing the degree of budding. The stars in the data correspond to the images in (a) as indicated by the numbers. (c) Wettabilities, as defined in the text, as a function of the contact angle $\theta_e$.

Table of Contents graphic:

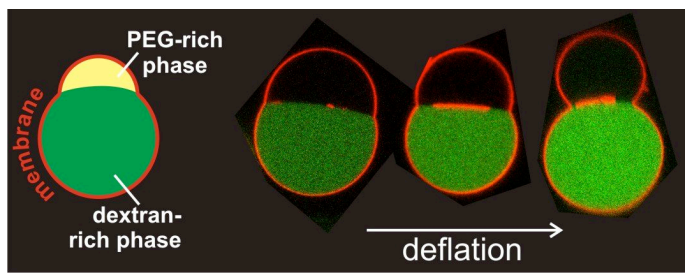

# Wetting-induced budding of vesicles in contact with several aqueous phases

## Supplementary Information

*Yanhong Li, Halim Kusumaatmaja, Reinhard Lipowsky, Rumiana Dimova**

Max Planck Institute of Colloids and Interfaces, Science Park Golm, 14424 Potsdam, Germany

* Address correspondence to: dimova@mpikg.mpg.de

**Materials:** Poly(ethylene glycol) or PEG (average molecular weight 8000 g/mol) and dextran from Leuconostoc mesenteroides (molecular weight between 400 kDa and 500 kDa) were purchased from Sigma-Aldrich. The polydispersity, measured with gel permeation chromatography, was 1.11 for PEG and 1.83 for dextran. Sucrose was purchased from Fluka. Lipids including 1,2-dioleoyl-sn-glycero-3-phosphatidylcholine (DOPC), and 1,2-dipalmitoyl-sn-glycero-3-phosphatidylethanolamine-N-(lissamine rhodamine B sulfonyl) as chloroform solutions and galbeta1-3galnacbeta1-4(neuacalpha2-3)galbeta1-4glcbeta1-1'-cer ($G_{M1}$ Ganglioside) as powder were purchased from Avanti Polar Lipids. In some occasions, instead of $G_{M1}$ we used membranes containing 4% dioleoylphosphatidylethanolamine-N-[methoxy (polyethylene glycol)-2000] (ammonium salt) (DOPE-PEG2000); data not shown. $G_{M1}$ and DOPE-PEG2000 were added in order to tune the interaction of the membrane with one of the two aqueous phases enclosed in the vesicles. Since the headgroup of $G_{M1}$ consists of sugar moieties, it is expected to induce stronger interactions of the membrane with the dextran-rich phase, while DOPE-PEG2000 is expected to interact stronger with the PEG-rich phase.

**Giant vesicle preparation using electroformation:** Lipid stock solution in chloroform (25~30 μl, 2 mg/ml) was spread on conductive glass substrates coated with indium tin oxide (ITO). The lipid films were dried in a vacuum desiccator for at least 3 hours. A rectangular Teflon frame of thickness 1.6 mm served as a chamber spacer between two opposing glass substrates. The chamber was sealed with grease. The coated ITO surfaces acted as electrodes. Approximately 2 ml polymer solution was injected into the chamber through a 0.22 μm filter. The chamber, the filter, the injection syringe, the needle and the polymer solution were preheated at 60 °C in an oven. The filled chamber was placed in the oven at 60 °C, and immediately afterwards, an AC field of 1.5 ~ 2.2 V (peak-to-peak, according to the resistance of the ITO glasses) and 10 Hz was applied using a function generator (Agilent 33220A 20MHz function/arbitrary waveform generator). The electroformation continued for 2 or 3 hours. The vesicles were prepared at 60 °C to ensure homogeneity of the polymer solution. After the formation, the chamber was cooled to room temperature and the vesicle solution was transferred into a small tube.

**Phase diagram of the polymer aqueous solution:** The binodal of the PEG and dextran aqueous solution at room temperature was determined by cloud-point titration for each polymer concentration combination; see Fig. S1. At first, concentrated PEG and dextran aqueous stock solutions were prepared. The mass of a small well-sealed vial with a stirring bar inside was measured with a balance (Mettler AT261 DeltaRange). Then, a certain amount of the dextran stock solution and water was injected inside, and the added weight measured. The PEG stock solution was injected drop-wise under stirring until the solution in the vial became turbid and the added weight was measured again. The same procedure was repeated for several different initial dextran concentrations. The cloud point titration was done at 23°C.

Note that the molecular weight and the polydispersity of the polymers can influence the binodal. In a similar fashion, the vesicle behavior is very sensitive to the concentration of the polymer solutions.

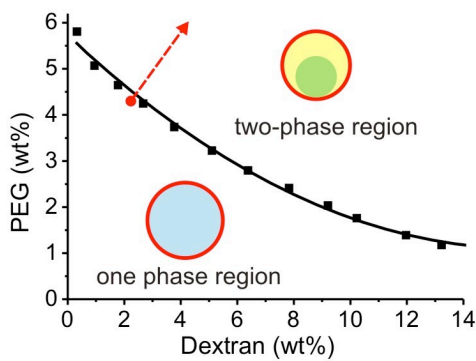

Figure S1. Phase diagram of the PEG-dextran polymer solution at 23ºC. The black squares indicate the binodal measured at 23ºC (the solid line is a guide to the eye). Below the binodal the polymer solution is homogeneous; above the binodal it phase separates. The red dashed line indicates the experimental trajectory of the deflation of vesicles initially loaded with the polymer solution (red dot: 4.05% PEG, 2.22% dextran). The insets schematically illustrate the vesicle membrane (red) enclosing the homogeneous solution (blue) or the two separated phases, dextran-rich (green) and PEG-rich (yellow).

**Batch measurements on deflated vesicles:** We performed deflation measurements on populations of vesicles, which are initially spherical. After deflation the vesicles change their morphology to a budded state and the intrinsic contact angle for all vesicles is the same. The morphological change after this single-step deflation was examined on altogether 64 vesicles. Since the preparation procedure of electroformation produces vesicles with different initial excess areas (or membrane tensions) and different volumes, each vesicle in the batch reaches a different degree of budding after the deflation step. Some examples are given in Fig. S2.

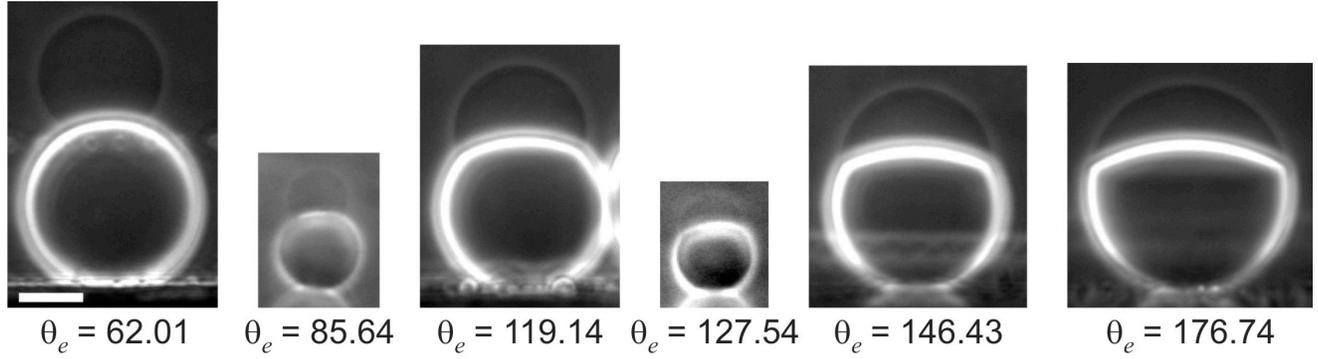

$\theta_e = 62.01$    $\theta_e = 85.64$    $\theta_e = 119.14$    $\theta_e = 127.54$    $\theta_e = 146.43$    $\theta_e = 176.74$

Figure S2. Phase contrast images (side view observation) of different vesicles after a single-step deflation. The lower droplet consists of the heavier dextran-rich phase, the upper droplet of the lighter PEG-rich phase. All images have the same magnification. The scale bar in the first snapshot corresponds to 20 $\mu$m.

**Force balance and intrinsic contact angle for inward budding:** In the main text, the outward budding process was analyzed using force balance arguments in analogy to Neumann's triangle for capillary surfaces. In the following, we will discuss the process of inward budding. In the experimental system, see Fig. 1(g, h), both the internal medium and the external continuous media (black regions) are PEG-rich phases. For clarity, we denote the inner medium by $i$, and the external one by $p$; the dextran-rich phase is denoted by $d$. Furthermore, we define the three contact angles, $\theta_i$, $\theta_p$, and $\theta_d$, see Fig. S3a, obtained from the microscopy images. At the contact line, the membrane tensions $\hat{\Sigma}_{pi}$ and $\hat{\Sigma}_{di}$, as well as the interfacial tension $\Sigma_{pd}$ between the external liquid phases must balance. We then obtain the relations

$$\hat{\Sigma}_{di} + \hat{\Sigma}_{pi} \cos\theta_i + \Sigma_{pd} \cos\theta_d = 0,$$
$$\hat{\Sigma}_{di} \cos\theta_i + \hat{\Sigma}_{pi} + \Sigma_{pd} \cos\theta_p = 0, \quad \text{(S1)}$$
$$\hat{\Sigma}_{di} \cos\theta_d + \hat{\Sigma}_{pi} \cos\theta_p + \Sigma_{pd} = 0,$$

with $\theta_i + \theta_p + \theta_d = 2\pi$. Furthermore, we can define the intrinsic contact angle of the external PEG-rich phase with the membrane in an analogous way to Ref. [1] for outward budding. As shown in Fig. S3b, this angle satisfies the simple relation

$$\cos\theta_{in} = \frac{\hat{\Sigma}_{di} - \hat{\Sigma}_{pi}}{\Sigma_{pd}}. \tag{S2}$$

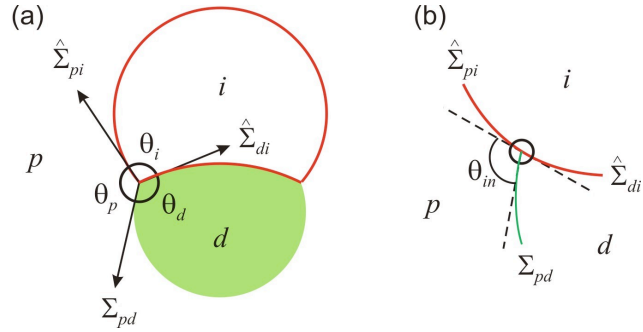

Figure S3. (a) Cross section of vesicles enclosing one droplet of interior phase $i$ in contact with one exterior droplet of dextran-rich phase $d$ (green) and the exterior bulk phase $p$. The thick red line represents the membrane. (b) Schematic illustration of the intrinsic contact angle of the external PEG-rich phase with the membrane. The three contact angles $\theta_i$, $\theta_d$, and $\theta_p$ and the three tensions $\hat{\Sigma}_{pi}$, $\Sigma_{pd}$, and $\hat{\Sigma}_{di}$ satisfy the three balance relations (S1).

**Interfacial tension measurements:** The bulk polymer solutions were prepared by mixing certain amount of PEG stock solution, dextran stock solution and water in 50 ml separatory funnels, respectively. The weight of each component was measured with a balance. The weight ratio of dextran and PEG was fixed to 0.55, as the one in the homogeneous polymer solution (2.22 wt% dextran and 4.05 wt% PEG) in all samples. The total weight in each funnel was around 60 g solution. The solutions were shaken by hand to make the polymers mix well. They were left at 24.2 °C (room temperature with small fluctuations) for 4 to 5 days. Afterwards, the two phases were separated. The dextran-rich phase was taken out through the lower outlet, and the PEG-rich phase through the upper outlet

The interfacial tensions of the coexisting phases were measured at 24.2 °C using a spinning drop tensiometer (SITE100SQ, Krüss). The density differences of the coexisting phases, which were needed to calculate the interfacial tension, were measured at 24.2 °C with a density meter (DMA5000, Anton Paar).

**Vesicle aspiration with micropipettes:** The freshly formed vesicles containing 4.05 wt% PEG and 2.22 wt% dextran were diluted in the isotonic solution, and pre-deflated by adding the hypertonic solution (3.92 wt% PEG, 2.14 wt% dextran and 3.27 wt% sucrose) into the external medium stepwise. The increment in the osmolarity ratio $r$ (see the main text for definition) was approximately 6.5% for each step, and the time interval between two steps was at least 15 minutes to avoid budding during deflation. At the end, the system was left over night to equilibrate. The deflated vesicles were carefully transferred into the working chamber. The micropipettes were pulled by a micropipette puller (P-97, Sutter Instrument Co.), and the tips shaped with a micro forge (MF-900, Narishige). The pipette inner diameters were typically ~20 μm. A pipette was inserted into the chamber from the top. To eliminate adhesion, the tip was pre-coated with lipids by breaking several vesicles. The opened side of the chamber was sealed with high viscous grease to avoid evaporation. Aspiration was realized by means of a hydrostatic pressure system with a motorized vertical stage (M-531.PD, Physik Instrumente). The system was left to equilibrate for about 3 minutes after each consecutive pressure change. Vesicle aspiration was observed from the side. All experiments were performed at room temperature.

The pipette tip was in contact with the membrane enclosing the PEG-rich phase, therefore the tension controlled via the pipette is $\hat{\Sigma}_{pe}$. It can be calculated from the following equation [2].

$$\hat{\Sigma}_{pe} = \frac{PR_{pip}}{2(1 - R_{pip}/R_p)} \quad (S3)$$

where $P$ is the suction pressure, $R_p$ is the radius of the PEG-rich phase, and $R_{pip}$ is the radius of the spherical part of the vesicle inside the micropipette.

The change in the contact area between the dextran- and the PEG-rich phases corresponding to the measurement shown in Fig. 3 is given in Fig. S4. Similar measurements were performed on altogether four vesicles.

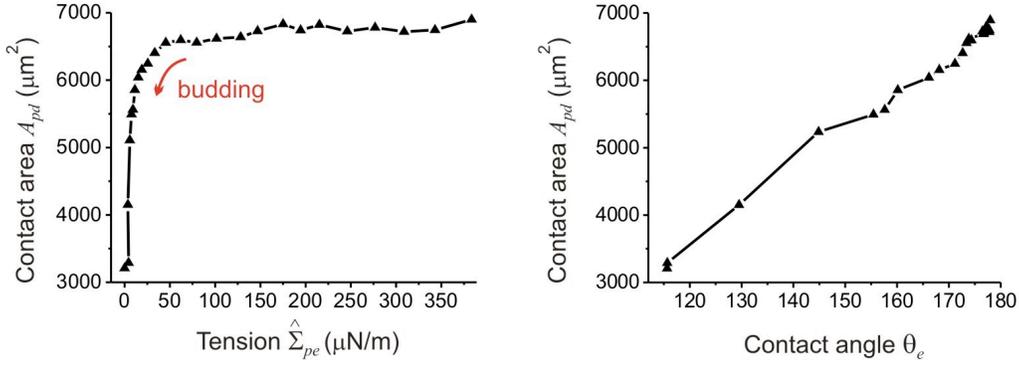

Figure S4. Change in the contact area $A_{pd}$ for the vesicle in Fig. 3 in the main text as a function of the tension $\hat{\Sigma}_{pe}$ applied with a micropipette (left panel) and the same data as a function of the contact angle $\theta_e$ (right panel).

**Derivation of Eq. (3) in the main text:** In the limit where $\theta_e$ approaches $\pi$, we can write

$$\theta_e = \pi - \delta \tag{S4}$$

where $\delta$ is a small parameter, $\delta \to 0$. Substituting (S4) into the relation $\theta_p + \theta_d + \theta_e = 2\pi$, we have

$$\theta_d = \pi + \delta - \theta_p \ . \tag{S5}$$

Thus,

$$\frac{\sin\theta_p - \sin\theta_d}{\sin\theta_e} = \frac{\sin\theta_p - \sin(\theta_p - \delta)}{\sin\delta} \approx \cos\theta_p + \sin\theta_p \tan\left(\frac{\delta}{2}\right) \approx \cos\theta_p + \frac{\delta}{2}\sin\theta_p. \tag{S6}$$

**Calculated vesicle shapes:** In the main text, the budding process was analyzed using force balance arguments. The same conclusions can be obtained from a systematic theory that includes the membrane bending energy [1]. The experimental input parameters are the vesicle volume and area as well as the volume ratio of the PEG-rich and dextran-rich phases, which are estimated from the microscopy images. Given these constraints, one obtains a family of solutions corresponding to different local energy minima. Thus, the shape of the vesicle cannot be uniquely computed. However, identifying the experimentally measurable property $(\hat{\Sigma}_{de} - \hat{\Sigma}_{pe})/\Sigma_{pd}$ with the cosine of the intrinsic contact angle as in Eq. (4) in the main text, the vesicle shape can be uniquely predicted without any fitting parameter. The computed shapes are in excellent agreement with the vesicle images; see Fig. S5.

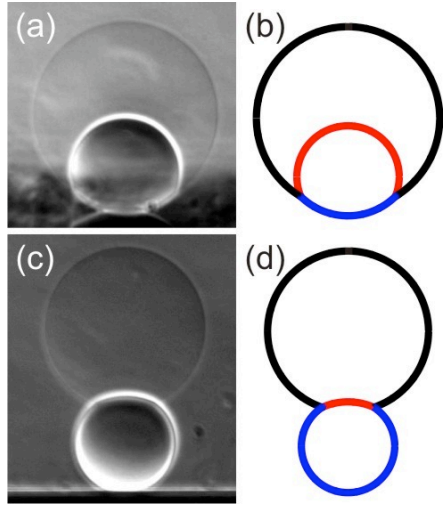

Figure S5. Comparison between vesicle images and theoretically calculated shapes. (a, c) Phase contrast images corresponding to the snapshots given in Fig. 1c,f in the main text. (b, d) computed cross-sections based on the theory discussed in [1].

**Saturation of the tension $\hat{\Sigma}_{pe}$ at high osmolarity ratios:** The tension $\hat{\Sigma}_{pe}$ saturates for high polymer concentration; see Fig. 2c in the main text. Because $\hat{\Sigma}_{pe}$ is governed by the membrane spontaneous curvature [3], we expect that the latter reaches saturation as well. One contribution to the spontaneous curvature arises from the different composition of the exterior solution and the two aqueous phases within the vesicles. For the systems studied here, the exterior solution contained PEG, dextran, and sucrose, whereas the interior PEG-rich and dextran-rich phases contained primarily PEG and dextran, respectively. The associated compositional asymmetry across the *pe* and *de* membranes leads to spontaneous or prefered curvatures $m_{pe}$ and $m_{de}$ of the two membrane segments. More precisely, these spontaneous curvatures reflect the different interactions between the membranes and the three molecular species [4,5].

If PEG molecules were adsorbed onto the membranes, for example, two layers of PEG would be formed on the two sides of the membranes, which would then prefer to curve away from the denser layer. In general, one expects to see competitive adsorption of several species such as PEG and dextran. In the latter case, the spontaneous curvature would saturate for high polymer concentrations if the interior and exterior membrane surfaces became completely covered with PEG and dextran and the corresponding adsorption layers differed in their composition.

As shown in Fig. S6, the PEG-rich phase contains essentially no dextran for large polymer concentrations. Therefore, in this concentration regime, the adsorption layer at the interior surface of the *pe* membrane would consist only of PEG whereas the adsorption layer at the exterior surface of the *pe* membrane would consist of both PEG and dextran. It remains to be seen if such a molecular mechanism can be corroborated by further experimental studies.

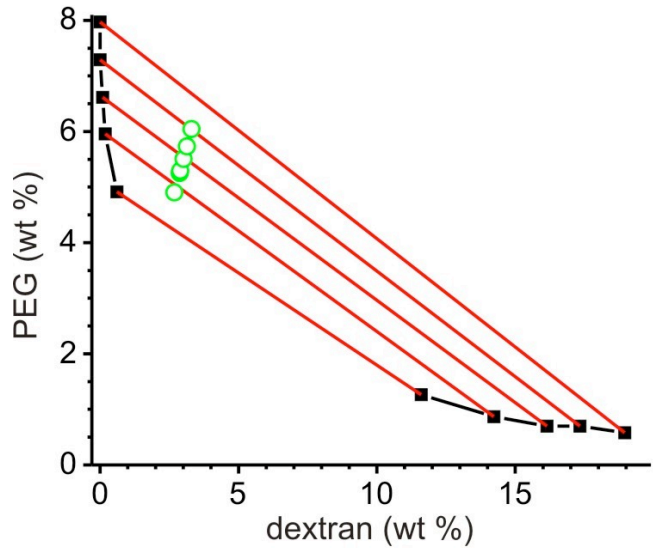

Figure S6. Coexistence curve and tie lines of dextran and PEG aqueous solutions as measured with gel permeation chromatography at 23 °C. The black curves are a guide to the eyes. The red lines illustrate the tie lines. At high polymer concentrations, no dextran is detected in the PEG-rich phase. The green open circles indicate the composition of polymer solutions explored in this work.